\documentclass[a4paper]{jpconf}
\usepackage{graphicx}
\usepackage{color}
\begin{document}
\title{Space-time breather solution \\ for nonlinear Klein-Gordon equations}

\author{Yasuhiro Takei$^1$, Yoritaka Iwata$^{2,*}$}

\address{$^1$Mizuho Information $\&$ Research Institute, Tokyo, Japan \\ 
$^2$Kansai University, Osaka, Japan}

\ead{$^*$iwata$\_$phys@08.alumni.u-tokyo.ac.jp}

\begin{abstract}
Klein-Gordon equations describe the dynamics of waves/particles in sub-atomic scales.
For nonlinear Klein-Gordon equations, their breather solutions are usually known as time periodic solutions with the vanishing spatial-boundary condition.
The existence of breather solution is known for the Sine-Gordon equations, while the Sine-Gordon equations are also known as the soliton equation.
The breather solutions is a certain kind of time periodic solutions that are not only play an essential role in the bridging path to the chaotic dynamics, but provide multi-dimensional closed loops inside phase space. 
In this paper, based on the high-precision numerical scheme, the appearance of breather mode is studied for nonlinear Klein-Gordon equations with periodic boundary condition.
The spatial periodic boundary condition is imposed, so that the breathing-type solution in our scope is periodic with respect both to time and space.
In conclusion, the existence condition of space-time periodic solution is presented, and the compact manifolds inside the infinite-dimensional dynamical system is shown.  
The space-time breather solutions of Klein-Gordon equations can be a fundamental building block for the sub-atomic nonlinear dynamics.
\end{abstract}

\section{Introduction}
Let us consider one-dimensional wave equations.
The existence of breather solution \cite{93Denzler, 11Blank, 18Maier, 20Scheider} has been known for some nonlinear Klein-Gordon equations; e.g., for Sine-Gordon equations.
The breather mode is regarded as a kind oscillation.
Indeed, for one-dimensional cases, it behaves asymptotically damping for $|x| \to \infty$, and periodic for $t$. 
Such a periodic property leads to the oscillation.
On the other hand, the breather mode is not necessarily stable in most of nonlinear Klein-Gordon equations (for a textbook, see \cite{98bjorken}).

In this paper, utilizing the high-precision numerical code [Iwata-Takei],  for hyperbolic evolution equations, the breather solution is explored in the double-well type nonlinear Klein-Gordon equations.
By assuming the periodic boundary condition for the spatial direction $x$, here we are seeking a periodic solution for both time and space.
In this sense it is likely to be called the space-time breather solution.
On the other hand, since the model equation exactly correspond to the $\phi^4$-theory in the quatum field theory, the obtained solution is expected to bring about a new insight on the existence of nonzero mass states.

\begin{figure}[t]
\begin{center}
  \includegraphics[width=60mm,bb=9 9 358 434]{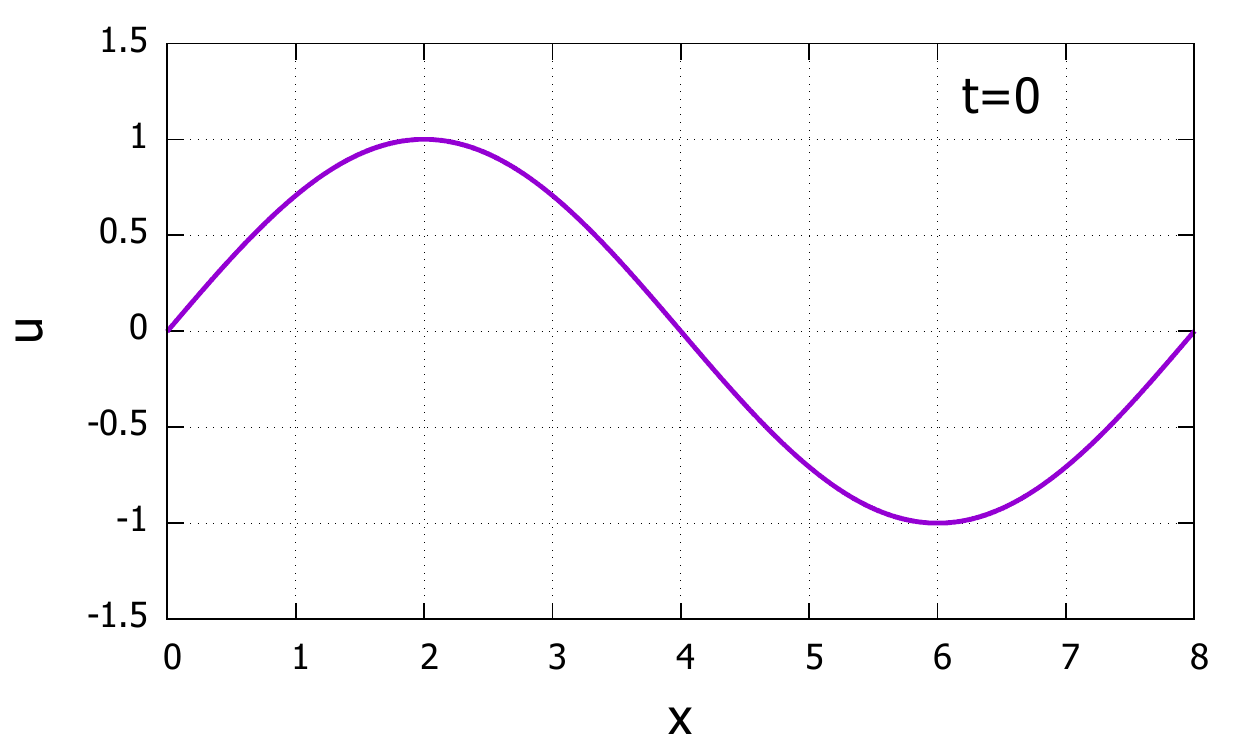}
  \quad 
  \includegraphics[width=60mm,bb=9 9 358 434]{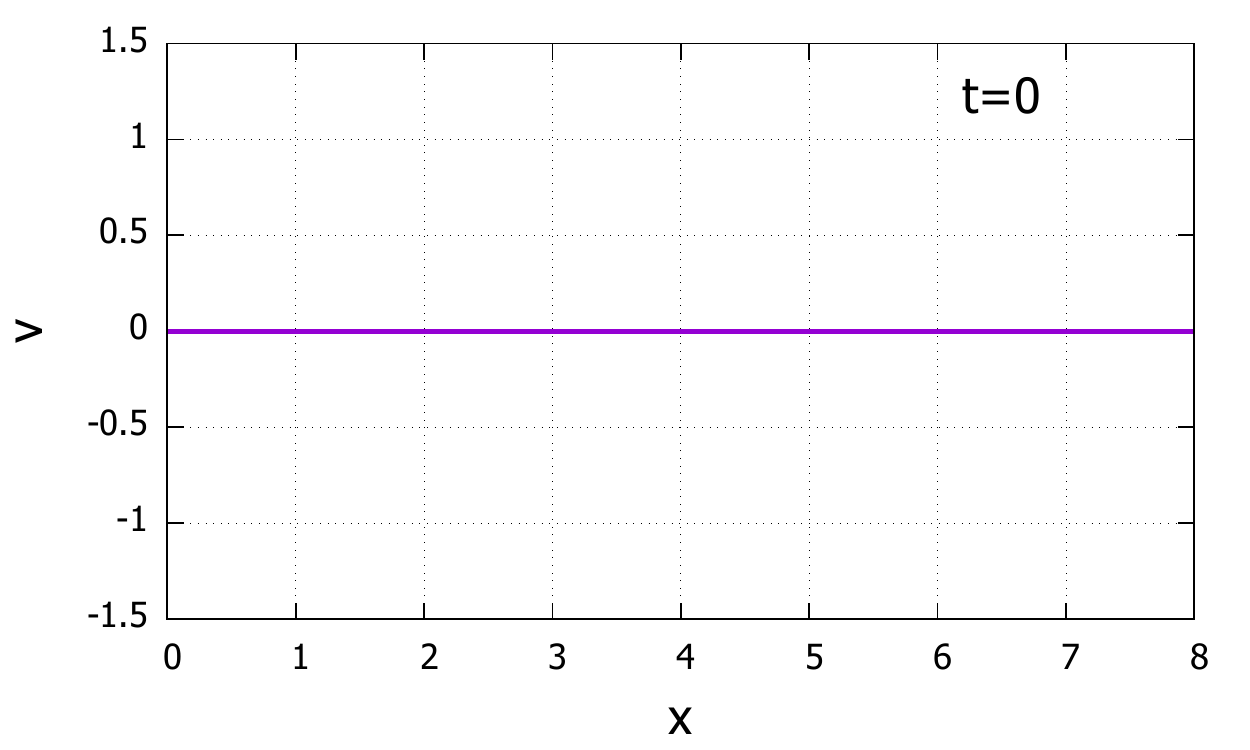} 
\caption{(Color online) Initial functions $f(x) = A \sin(\pi x /4)$ and $g(x)= 0$ with $A=1$, where the spatial range is fixed $[0,8]$.}
\label{fig:ini}
\end{center}
\end{figure}

\section{Mathematical model}
\subsection{Nonlinear Klein-Gordon equations with cubic nonlinearity}
Let $x \in [0,L]$ be a finite domain of space.
The positive evolution problem is considered ($t \ge 0$). 
Let $\partial/\partial t$ and $\partial/\partial x$ be denoted by $\partial_t$ and $\partial_x$ respectively.
We consider the nonlinear Klein-Gordon equation with the double-well type interaction, which is also known as $\phi^4$-theory.
\[
\label{eq:eq01}
\begin{array}{ll}
\ \partial_t^{2} u + \alpha \partial_x^{2} u +   (\beta u^2 - \mu) u  = 0, \vspace{6mm} \\
\ u(x, 0) = f(x),\ u(0, t) = u(L, t),  \vspace{6mm} \\
\ \partial_t u(x, 0) = g(x),\ \partial_t u(0, t) = \partial_t u(L, t).
\end{array}
\qquad {\rm (KG)}  
\]
where $\alpha$, $\beta$, and $\mu$ are real constants.
Since Eq. (KG) is solved by the Fourier transform, initial functions $f(x)$ and $g(x)$ are given as $L^2$-functions, and the periodic boundary condition is imposed for $x$-direction.
In the numerical calculations of this paper, the initial functions are fixed to $f(x) = A \sin(\pi x/4)$ and $g(x) = 0$ with $A>0$ (Fig.~\ref{fig:ini}).
By taking $v = \partial_t u$, the first equation of (KG) is written by
\begin{equation}
 \partial_t v + \alpha \partial_x^2 u  +  (\beta u^2 - \mu) u = 0. 
 \end{equation}
If  $\beta =0$ is satisfied, it is simply a linear Klein-Gordon equation in which $\mu$ means the square root of the mass. 
Otherwise if $\mu=0$ is satisfied (massless case), we see that Eq. (KG) is a generalization of Klein-Gordon equation with cubic nonlinearity (cf. $\phi^4$-theory in the context of quantum field theory (for a textbook, see \cite{65bjorken})).
This equation holds the symmetry breaking, which is known as the Higgs mechanism.

\subsection{Free-particle solutions}
The free-particle solutions are useful to identify the condition for the appearance of breather solution.
Let $\beta$ be a real constants satisfying $\beta > 0$.
By taking $(\beta u^2 - \mu) u =0$, the constant distributions (corresponding to three vacuums in the context of Higgs mechanism) follow:
\begin{equation} 
 u = 0, \quad  \pm \sqrt {\mu/\beta}    
\end{equation}
which trivially satisfy the initial and boundary value problem (KG). 
These solutions correspond to constant stationary solutions of (KG) without any interaction.
That is, $u = 0, \quad  \pm \sqrt {\mu/\beta}$ are regarded as free particle solutions, and Eq. (KG) always holds the free-particle solutions.
Needless to say, three solutions are degenerated to massless cases if $\mu\le 0$ is satisfied.
The stability of these three solutions depends both on the parameter settings.

Massless free-particle solutions (being obtained by setting $\mu = \beta = 0$) include other solutions than the constant solutions.
For example, let $A$ be a real number, 
\begin{equation}   \label{fwave}
u = A e ^{\pm i (kx - \omega t)}
\end{equation}
is a massless free-particle solution with the equality $k^2 = - \alpha \omega^2$, where initial functions should be $f(x) = A e ^{\pm i kx }$ and $g(x) = A \omega  e ^{\mp i (kx + \pi/2) }$ in this case.
These solutions correspond to typical solutions of (KG) in a limited setting. which are also useful to the mode analysis.
In this paper much attention is paid to the dynamics of nonlinear solutions. 
We will see that the dynamics of nonlinear solutions (interacting solutions) are highly affected by the free-particle solutions (non-interacting solutions).
In the theory of dynamical systems, there is a technical concept ``absorbing set'.
The constant solutions, which are also regarded as the stationary solutions, may or may not play a role of absorbing set (for the definition, see \cite{97temam}).

\begin{figure}[t]
\begin{center}
\hspace{-12mm}
  \includegraphics[width=48mm,bb=9 9 358 434]{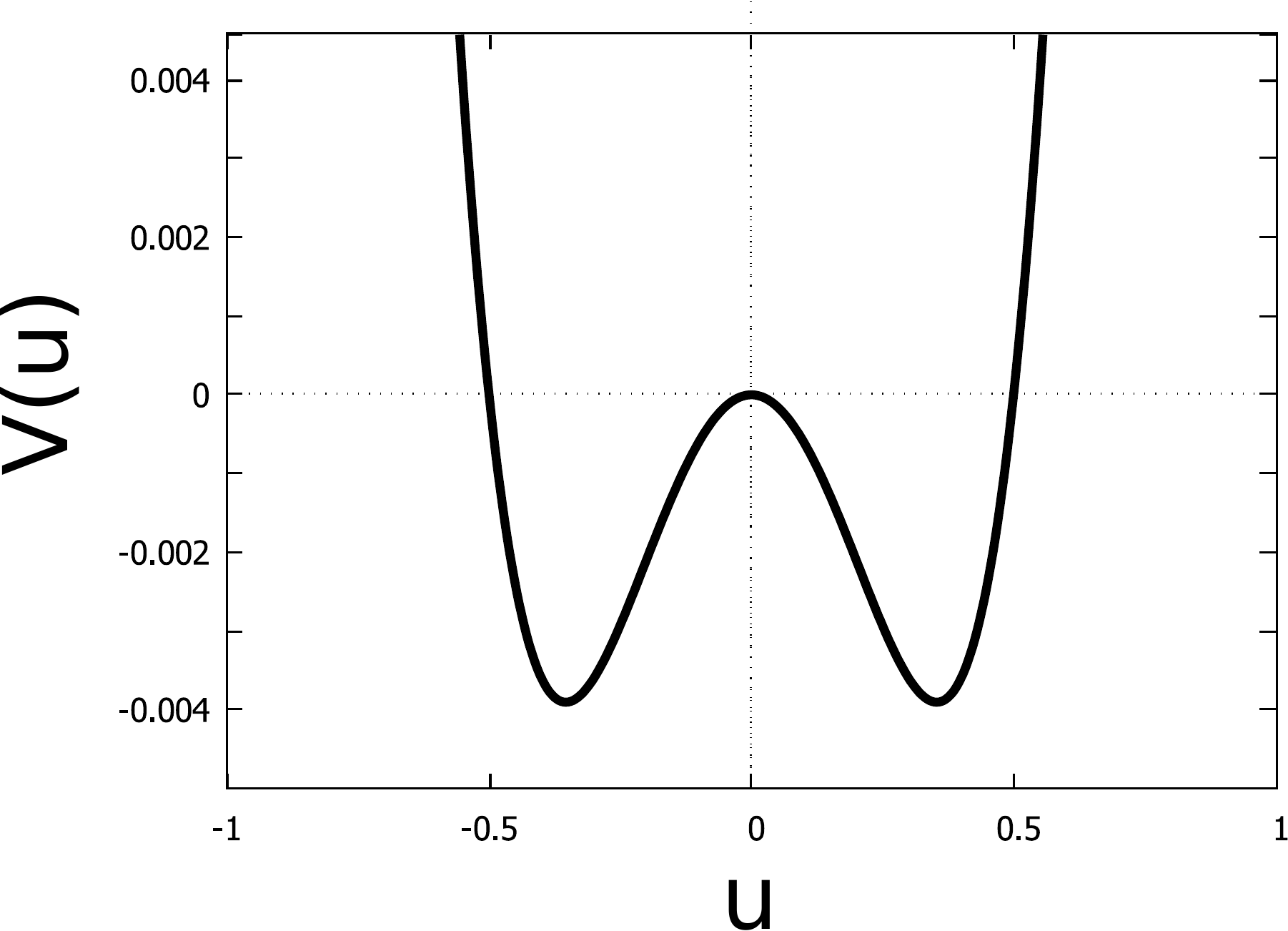}
  \caption{Higgs potential (\ref{higgspot}) in case of $(\mu,\beta)=(0.125,1)$.Two minimums correspond to $u= \pm \sqrt{0.125}$, and $u=0$ shows a maximum of the potential. }
\label{fig:higgs}
\end{center}
\end{figure}

\begin{figure}[tb]
\begin{center}
  \includegraphics[width=52mm,bb=9 9 358 234]{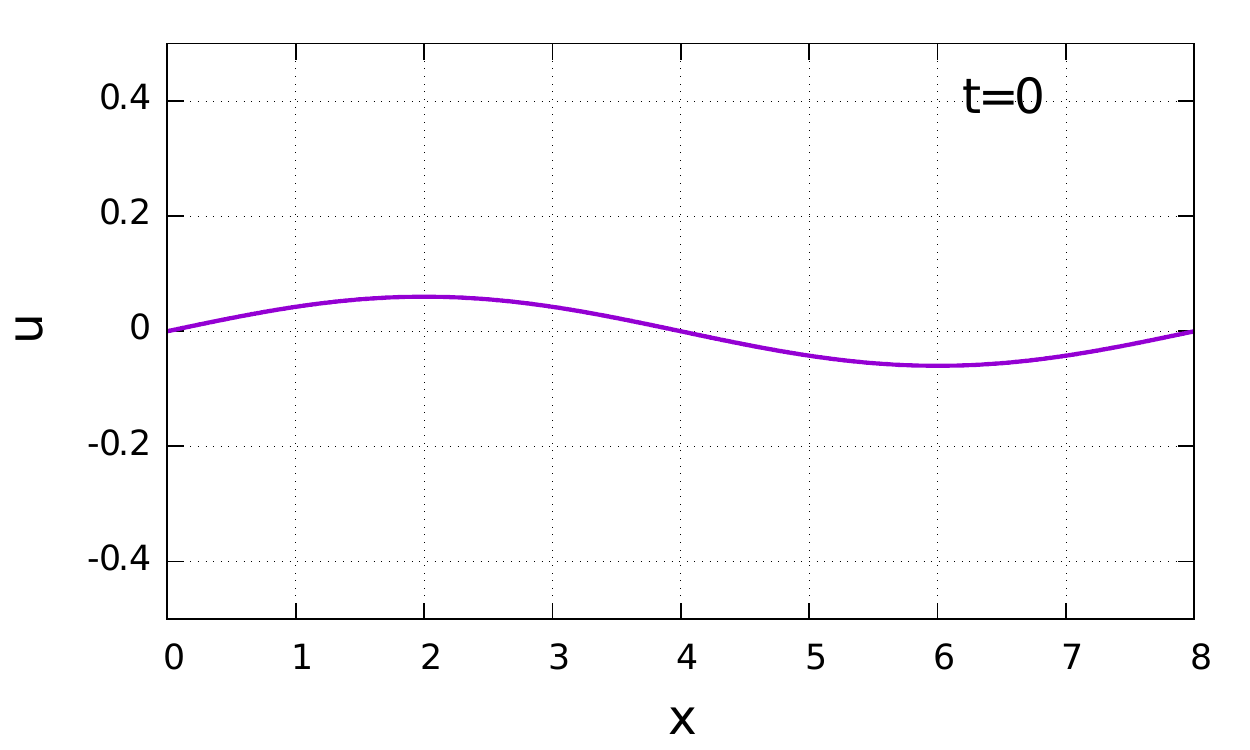} 
  \includegraphics[width=52mm,bb=9 9 358 234]{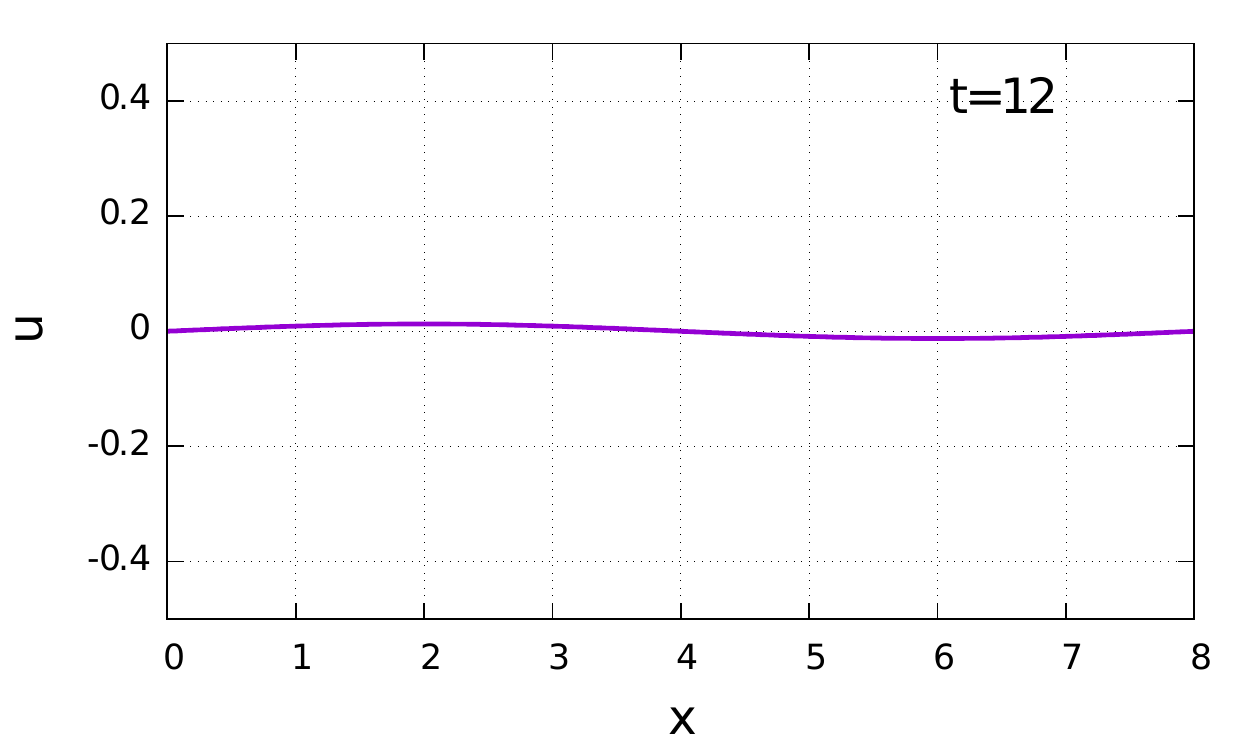} 
  \includegraphics[width=52mm,bb=9 9 358 234]{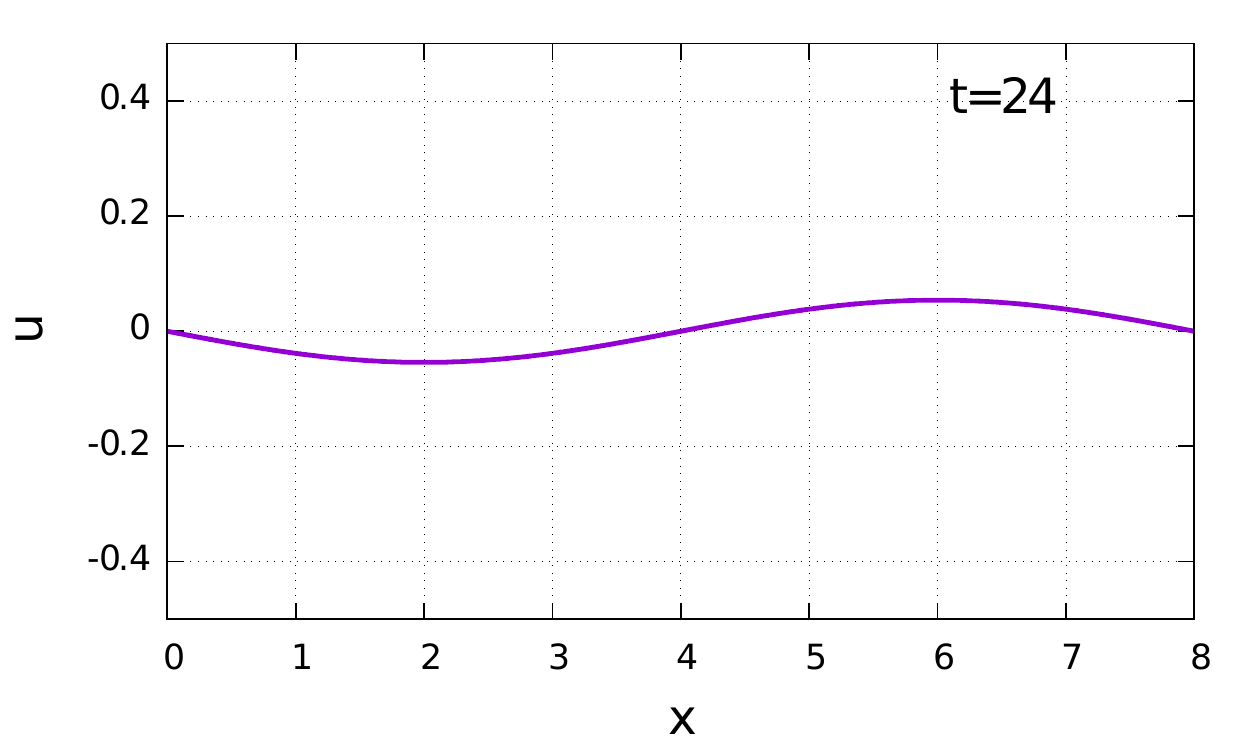}   \\
  \includegraphics[width=52mm,bb=9 9 358 234]{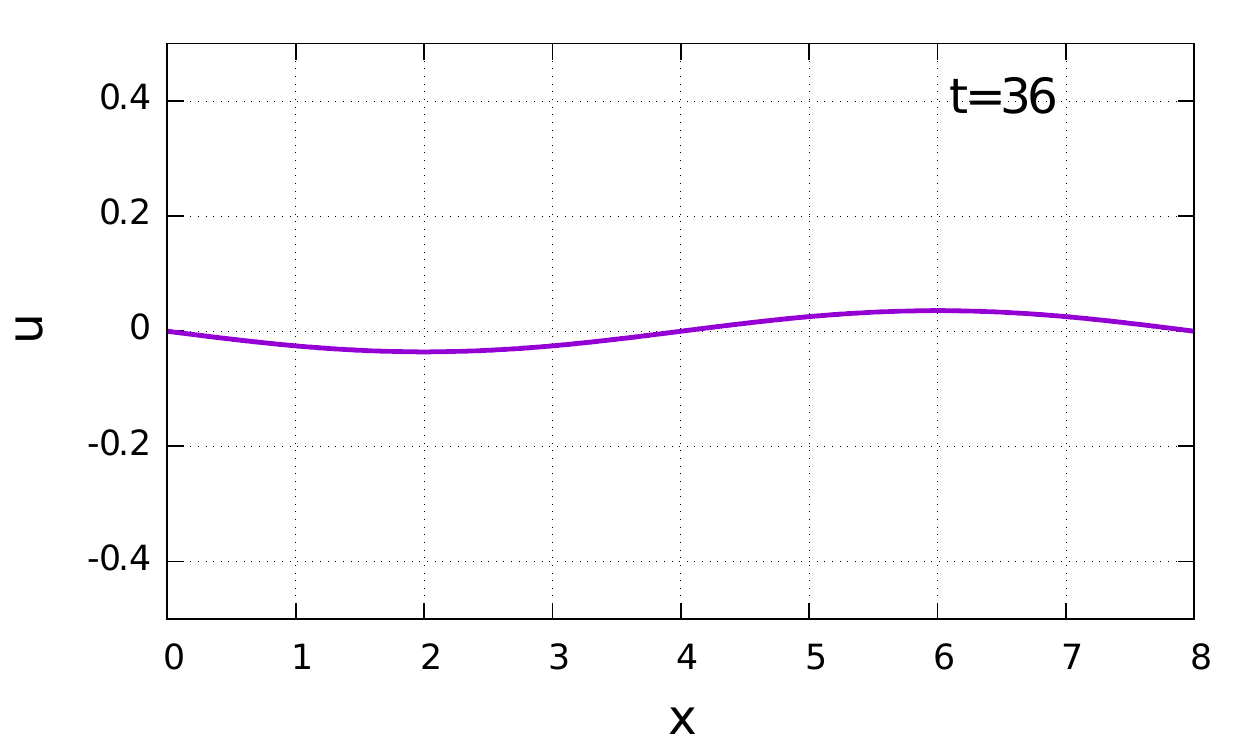} 
  \includegraphics[width=52mm,bb=9 9 358 234]{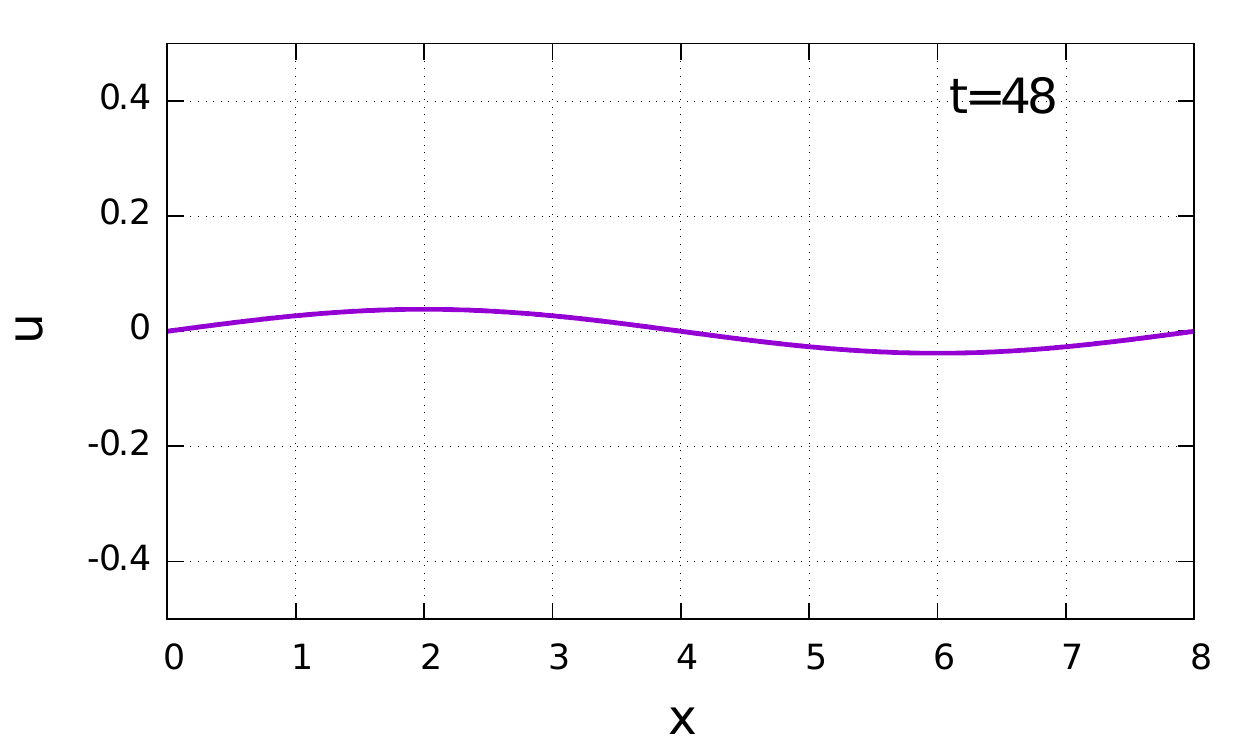} 
  \includegraphics[width=52mm,bb=9 9 358 234]{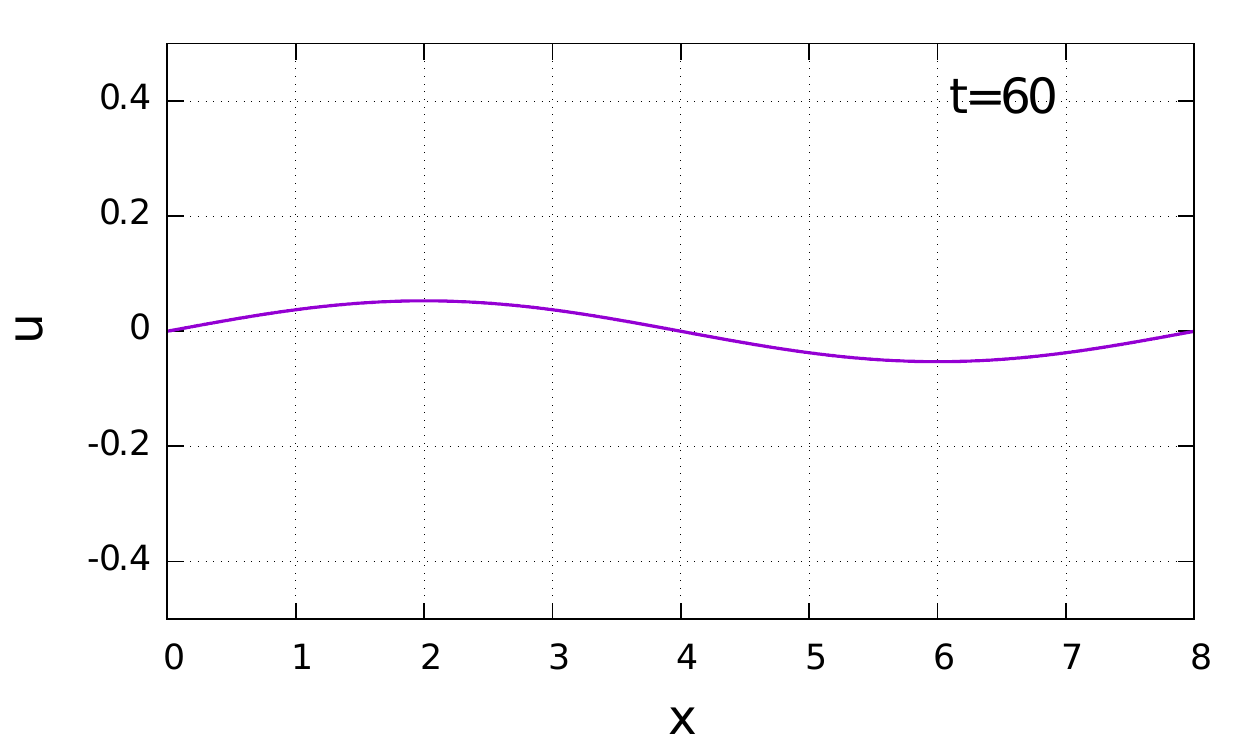} 
\caption{(Color online) $(A,\ \alpha)=(0.06,\ -0.220)$\ (Simply oscillating solution).
A part $u(x,t)$ with $x \le 4$ and another part $u(x,t)$ with $x > 4$ show the oscillations with its center $u=0$.}
\label{fig-o}
\end{center}
\end{figure}

\section{Theoretical estimates for the appearance of breather mode}
\subsection{Stability of constant distributions}
Following the usage in the quantum field theory, let us call
\begin{equation} \label{higgspot}
V(u) = \frac{1}{4} \beta u^4 - \frac{1}{2} \mu u^2
\end{equation}
the Higgs potential in case of the present inhomogeneous term $(\beta u^2 - \mu) u$.

\subsection{Profile of the breather solution}
\begin{figure}[tb]
\begin{center}
  \includegraphics[width=52mm,bb=9 9 358 234]{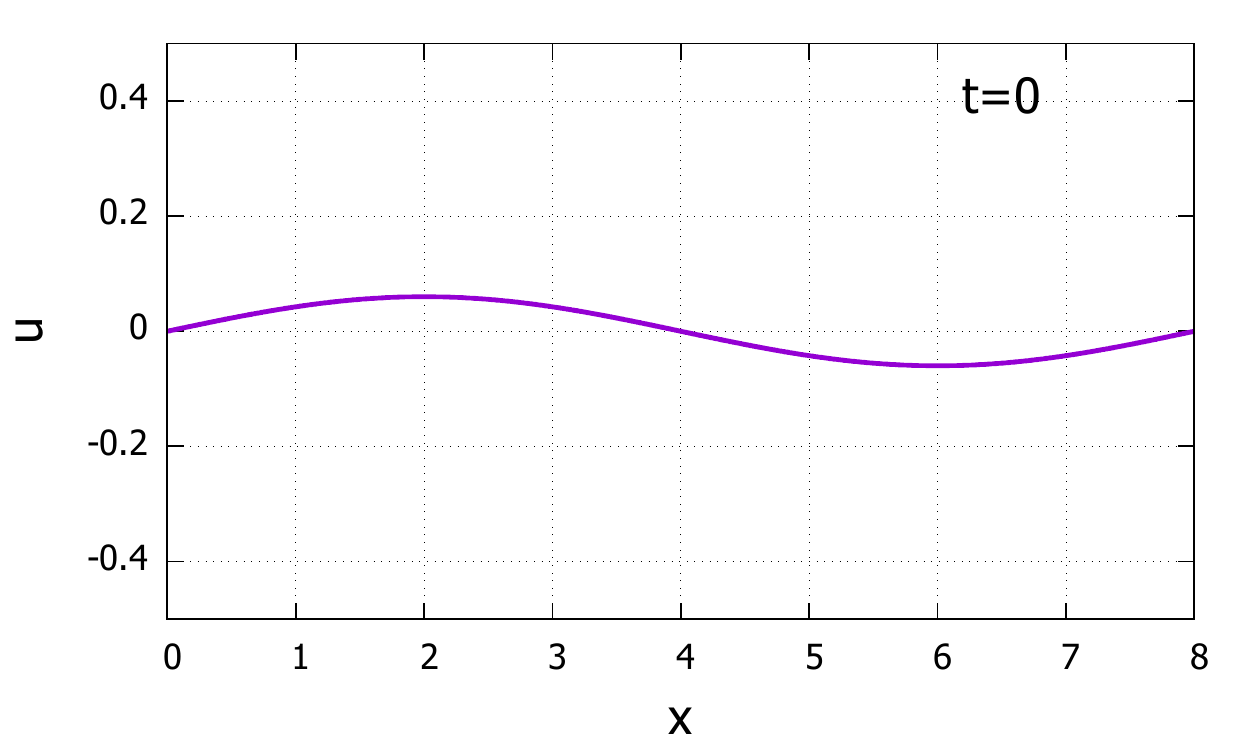} 
  \includegraphics[width=52mm,bb=9 9 358 234]{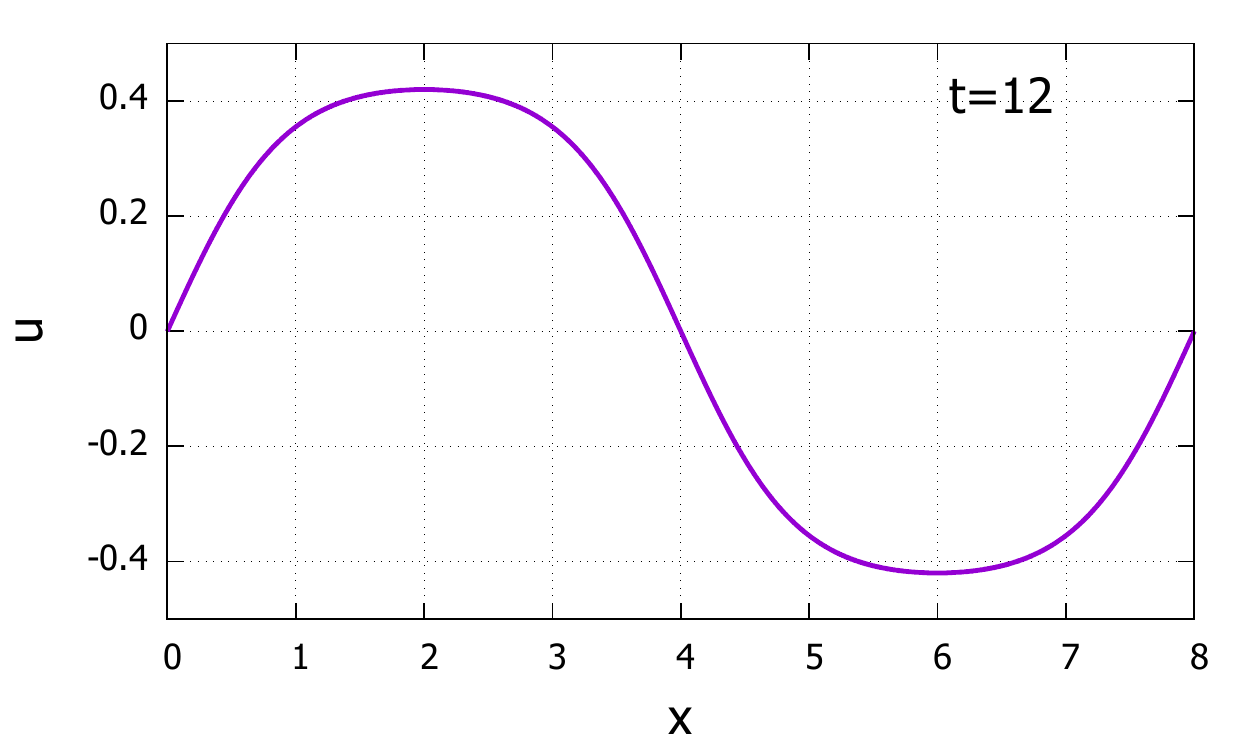} 
  \includegraphics[width=52mm,bb=9 9 358 234]{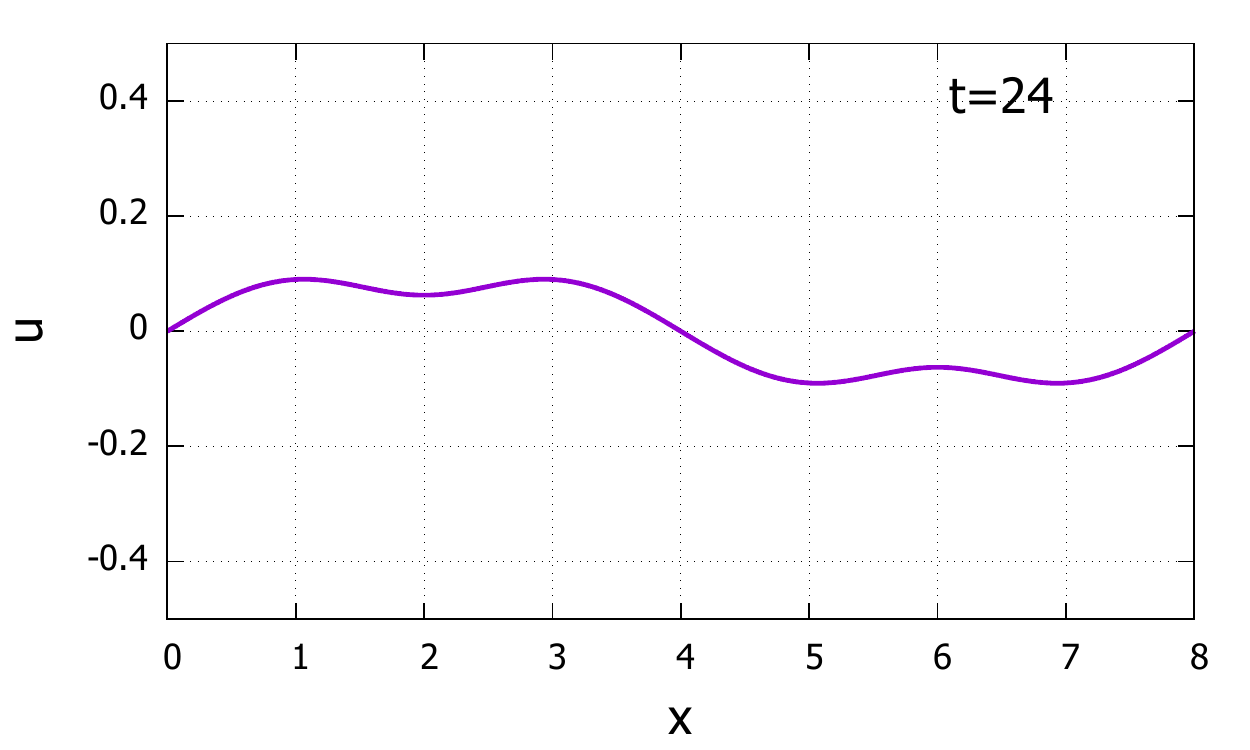}  \\
  \includegraphics[width=52mm,bb=9 9 358 234]{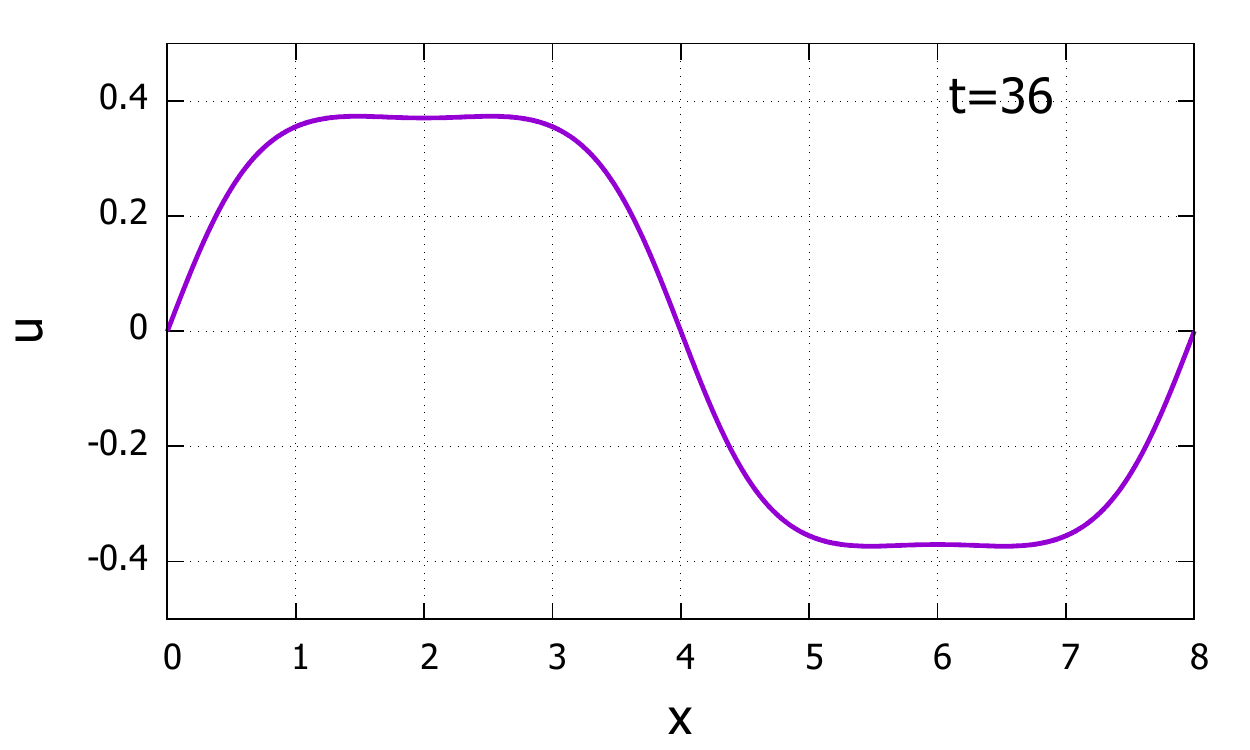} 
  \includegraphics[width=52mm,bb=9 9 358 234]{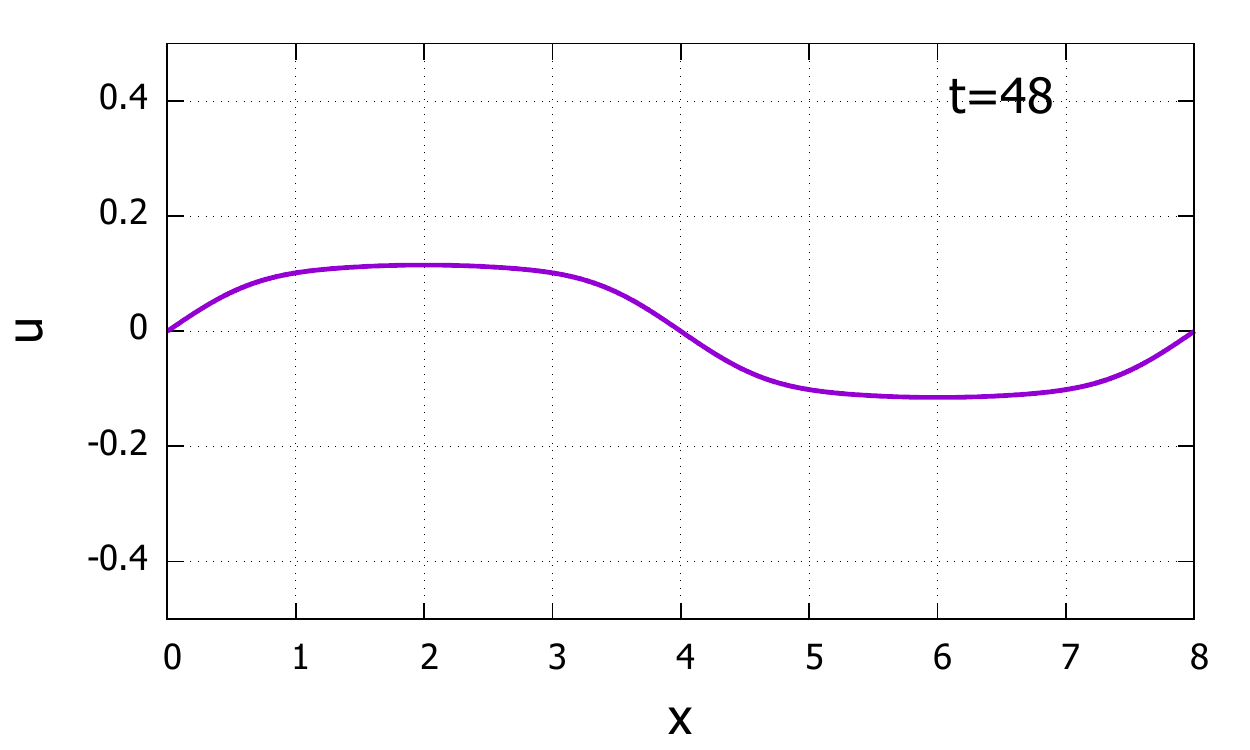} 
  \includegraphics[width=52mm,bb=9 9 358 234]{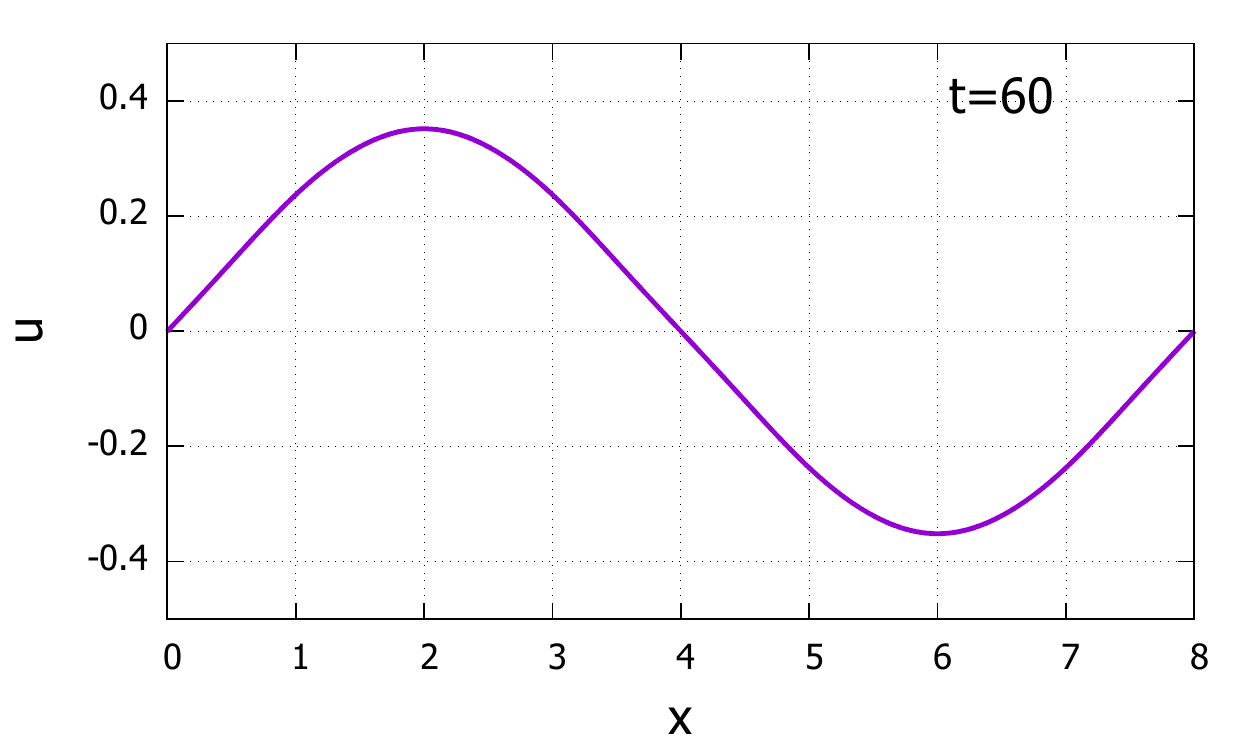} 
\caption{(Color online) $(A,\ \alpha)=(0.06,\ -0.075)$\ (Breather solution).
A part $u(x,t)$ with $x \le 4$ is confined in the positive region ($u(t,x)>0$) and a part $u(x,t)$ with $x > 4$ is in the negative region ($u(t,x)<0$).
}
\label{fig-b}
\end{center}
\end{figure}

The breather solution is the periodic solution for both $t$ and $x$, so that it is a kind of oscillation.
The breather solution can be distinguished from a simple oscillation (Fig.~\ref{fig-o}; for short, we call oscillation in the following) by the appearance of certain kinds of collectivity, where activated modes result in the resonance.
Here the breather solution (Fig.~\ref{fig-b}) includes the resonating large amplitude oscillation, which localized only in the positive or negative side of $u=0$.
This localization property is not satisfied by the simply-oscillating solution.
In this sense the terminology the breather mode makes sense in which many modes achieves the resonance.
Simply speaking, the breather solution is realized by the instability of constant solution $u=0$ and the stability of constant solutions $u = \pm \sqrt{\mu/\beta}$.

\subsection{Condition for the appearance of breather solution}
Here we obtain a guiding criterion in advance to a systematic calculation. 
Let a function $G(u)$ be defined by
\[
G(u) = -\alpha \partial_x^2 u -  (\beta u^2 - \mu) u,
\]
where $\alpha < 0$ and $\beta > 0$.
If we confine ourselves to a free-particle solution $u = (-Ai/2) (e ^{i (kx - \omega t)} -  e ^{-i (kx - \omega t)} )$ (see also Eq.~\ref{fwave}; in the following we call the formal solution) in the fully interacting cases, let $G(u)$ be replaced with
\[
{\tilde G}(u) = \alpha k^2 u -  (\beta u^2 - \mu) u
 =  \beta u \left( \frac{ \alpha k^2 + \mu}{\beta}  -   u^2 \right).
\]
This setting corresponds to the mode analysis for the stability.
Since $f(x)$ has the same form as $A \sin (kx) = - Ai (e^{ikx} - e^{-ikx})/2$, ${\tilde G}(u)$ is true at $t= 0$ at the least.
By considering the stationary condition $G(u)=0$, three roots are represented by
\[
u = 0,  \quad \pm \sqrt{ \frac{\alpha k^2+ \mu}{\beta} },
\]
where three real roots exist if $\mu > - \alpha k^2$,
three real roots are exactly the same if $\mu = - \alpha k^2$, and
one real root with two imaginary root  exist if  $\mu < - \alpha k^2$.
These solutions correspond to stationary solutions of (KG).
That is, the number of constant solution depend on the mass $\mu$, and more definitely on the activated spatial frequency $k$.

Under the condition $\mu > - \alpha k^2 $, let us limit ourselves to the constant distributions.
In case of ${ \bar u} =  \sqrt{ \frac{\alpha k^2+ \mu}{\beta} }$, the absorbing set is calculated by $(0, {\bar u_{\max}}]$ satisfying
\[
  \int_0^{\bar u_{\max}}  {\tilde G}(u) du 
=  \int_0^{\bar u_{\max}}   \beta u \left( \frac{ \alpha k^2 + \mu}{\beta}  -   u^2 \right) du 
=  \left[ \frac{1}{2} \left(  \alpha k^2 + \mu \right) u^2  -  \frac{\beta}{4} u^4  \right]_0^{\bar u_{\max}}  = 0,
\]
and consequently, the absorbing set is calculated to be
\begin{equation}   \label{cond_absorb} 
(0, {\bar u_{\max}}]  = \left( 0,  \sqrt { \frac{2( \alpha k^2 +   \mu) } {\beta} } \right] ,
\end{equation}
where
$
 (\alpha k^2+ \mu)/\beta  <  2  (\alpha k^2+ \mu)/\beta
$
is always satisfied, and ${\bar u} \in (0, {\bar u_{\max}}]$ is true.

\begin{figure}[tb]
  \begin{center}
  \hspace{-24mm}
  \includegraphics[width=50mm,bb=9 9 358 434]{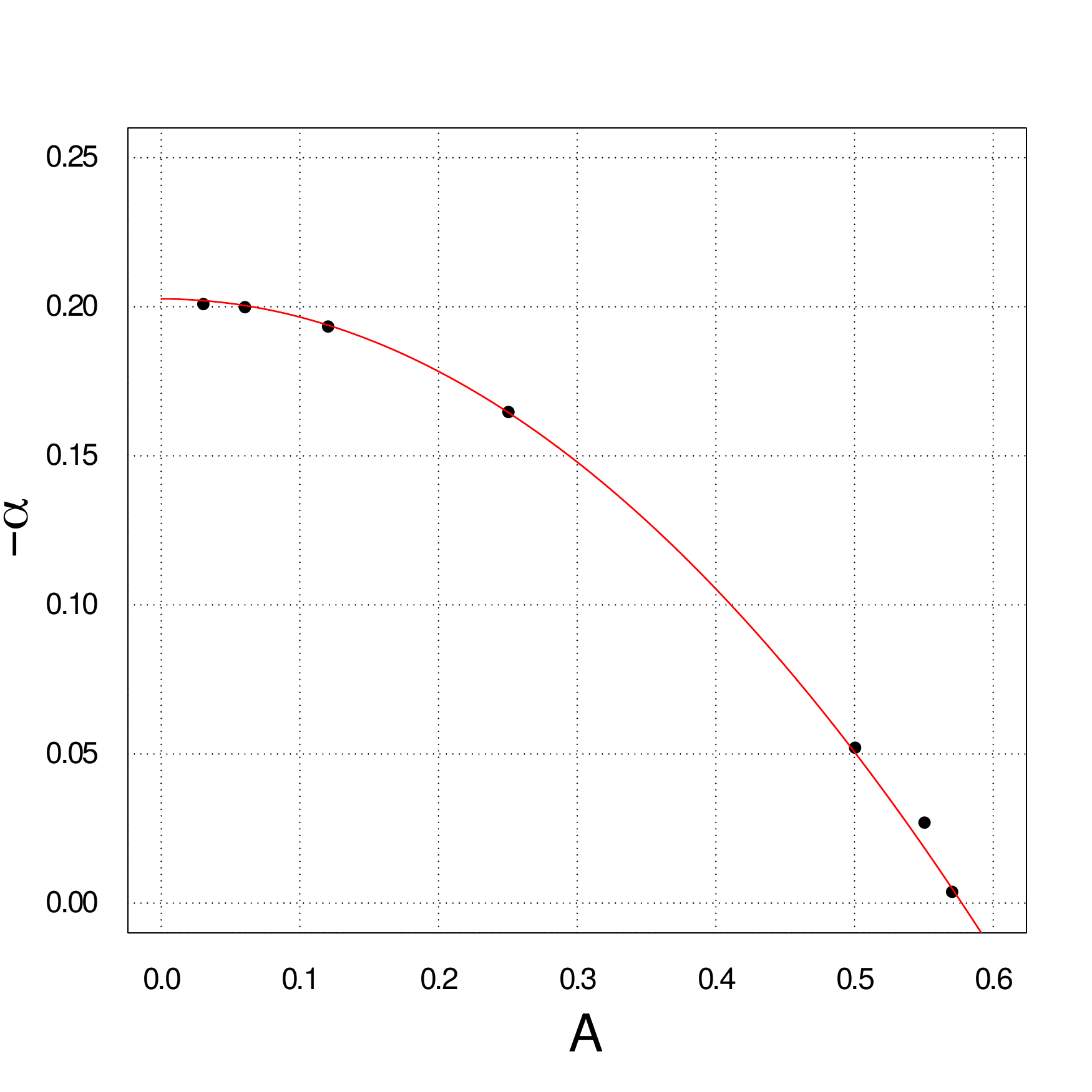}
  \caption{(Color online) The border of breather and oscillation solutions, where parameters are fixed to $(\mu, \beta, L) = (0.125,1,8)$.
    The validity of theoretical fomula (\ref{cond_bound}) is confirmed by the numerical results.
    The calculated values are shown by circles that show the breather solutions but some of those neighbor points cannot be the breather solution.
    Red curve is given by $ -\alpha = \frac{1}{k^2} \left(\mu-\frac{3}{8}\beta A^2 \right).$
    Note that we have confirmed by taking several time intervals that no breather solution can exist in the right-upper region of the graph.
}
\label{fig:res1}
\end{center}
\end{figure}

On the other hand, in terms of the curvature of spatial distribution, let us focus on the differential operator without approximation.
A part of spatial distribution is picked out by setting an interval $[{\bar x}_0, {\bar \xi}_0]$.
The differential operator satisfies
\[ 
  \int_{{\bar x}_0}^{\bar x_{\rm blc}}  \partial^2_x u ~ dx 
  =   \int_{\bar x_{\rm blc}}^{{\bar \xi}_0}  \partial^2_x u ~ dx, 
 \]
if the the 2nd derivative balances at $x = {\bar x}_{\rm blc}$.
Here the starting point ${\bar x}_0$ is assumed to satisfy $ \partial_x u ({\bar x}_0) = A$ (i.e. $A \cdot 1$ in the formal solution).
The ending point ${\bar \xi}_0$ is assumed to satisfy $ \partial_x u ({\bar \xi}_0) = 0$  (i.e. $i k A \cdot 0$ in the formal solution) that corresponds to the ending point.
Consequently the spatial infection point is expected be included in $[{{\bar x}_0}, {\bar \xi}_0]$.
The condition follows as
\[
 \partial_x u ({\bar x}_{\rm blc}) = \frac{1}{2} \partial_x u (x_0) = \frac{1}{2} A.
\]
If the formal solution $u = (-Ai/2) (e ^{i (kx - \omega t)} -  e ^{-i (kx - \omega t)} )$ is applied,
\begin{equation}   \label{cond_balance} 
 u  (\bar x_{\rm blc})  = \frac{\sqrt{3}}{2}A
  \end{equation}
is obtained, where the condition is checked for the discrete time $\omega t = \pi, 2 \pi, \cdots$ with respect to the mode analysis.
The resulting wave amplitude makes sense. 
Indeed, $\sqrt{3}A/2$ implies a wave amplitude indicating the balance of the 2nd spatial derivative.

By coupling two conditions (\ref{cond_absorb}) and (\ref{cond_balance}), we have the condition for the boundary of absorbing set with the balanced second order derivative.
If the condition
\begin{equation}   \label{cond_bound} 
 - \alpha    \le  \frac{1}{ k^2} \left( \mu  -  \frac{3}{8} \beta A^2 \right) 
\end{equation}
is satisfied, the positive constant solution ${\bar u} =  \sqrt{ \frac{\alpha k^2+ \mu}{\beta} } $ behaves as a local attractor with a absorbing set (\ref{cond_absorb}).
The similar analysis is valid for the negative constant solution ${ \bar u} = - \sqrt{ \frac{\alpha k^2+ \mu}{\beta}} $, while ${ \bar u} = 0$ is expected to repulse neighbor solutions at least for $\mu > - \alpha k^2$.
As a result, a statement to confirm in this paper is
\begin{itemize}
\item the breather solutions appear and survive at least for a while, if the initial function $f(x) = A \sin(\pi x /4)$ and $g(x)= 0$ is given to satisfy the condition (\ref{cond_bound})
\end{itemize}
where the further details are obtained as a life time formula.


\begin{figure}[tb]
  \begin{center}
  \hspace{-24mm}
  \includegraphics[width=50mm,bb=9 9 358 434]{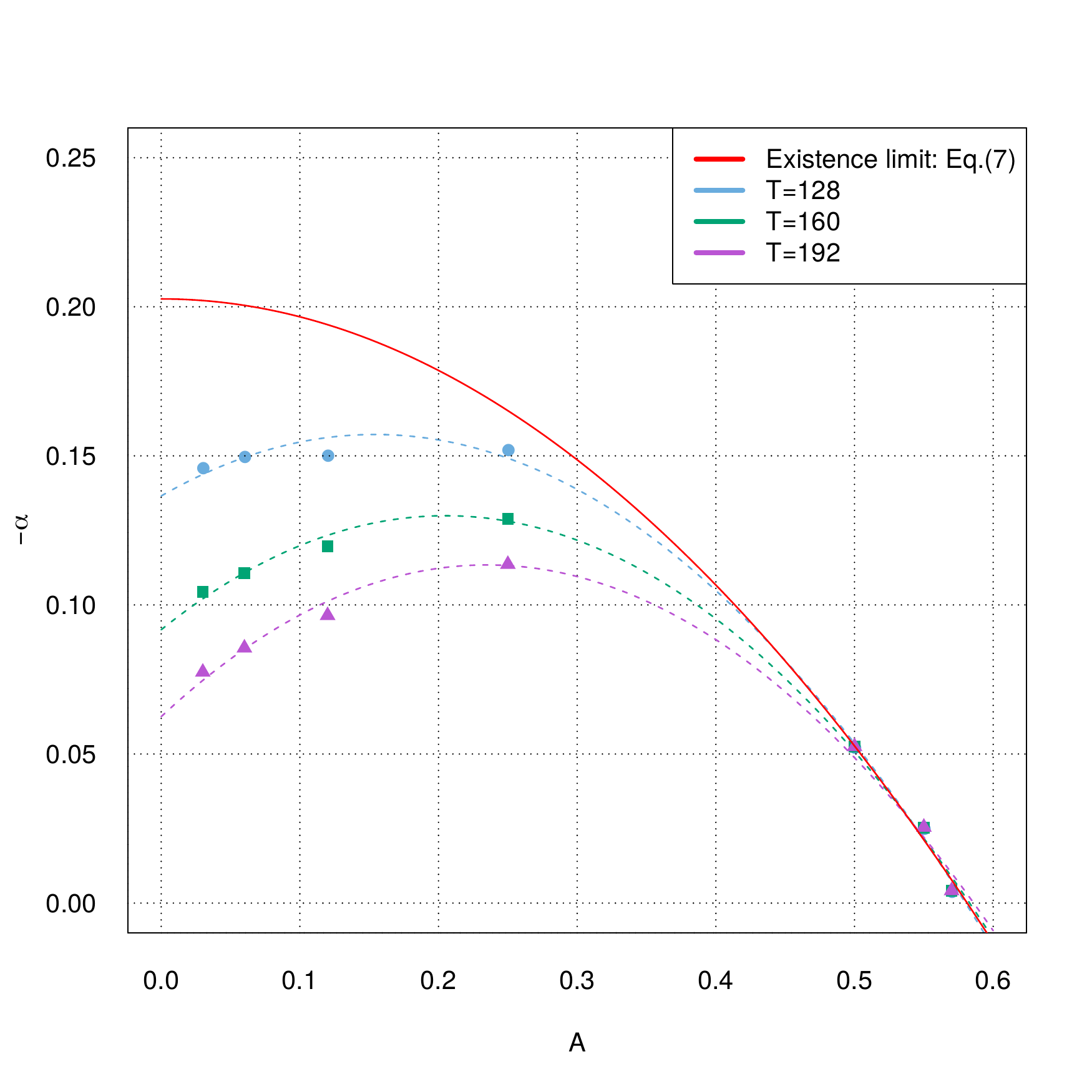}
  \caption{(Color online) Life time estimates for breather solutions, where parameters are fixed to $(\mu, \beta, L) = (0.125,1,8)$.
     Comparison between three different time intervals $[0,T],(T=128,160,192)$ is shown in order to find the law of lifetime.
    The calculated values are shown by circles, squares and triangles.
    They mean maximum of parameters $- \alpha > 0$ with the breathing oscillation modes to be stable by each time intervals $T=128, 160, 192$.
    Blue, green, purple curves are regression curves for the calculated values by each time intervals $T=128, 160, 192$.
    Red curve is given by $ -\alpha = \frac{1}{k^2} \left(\mu-\frac{3}{8}\beta A^2 \right).$
}
\label{fig:res2}
\end{center}
\end{figure}

\section{Numerical experiments}
\subsection{Settings}
The numerical calculations are carried out based on the high-precision numerical code using the Fourier spectral method \cite{20iwata-takei-proc01, 20iwata-takei-proc02}.
In the present version, the implicit third order Runge-Kutta method with two intermediate steps is utilized for the time direction, and spectral treatment is implemented for the space direction.
The spatial discretization is carried out based on the spectral method.
The solution is assumed to be expanded by the Fourier series, and terminated at the $2^{12}$ th term, which corresponds to the resolution for the spatial direction.
The time discretization used in the implicit calculation is fixed to $\Delta t = 2^{-13}$.
The size $L$ of the space is fixed to $[0, L] = [0,8]$, while the calculation ending time $T$ is flexible in order to identify the life time of breather solution.
$T =  64, 128, 160$, and $192$ are examined.

\subsection{Result}
The coefficients are fixed to $\mu = 0.1250$, $\beta = 1$.
As the initialization of this research, we begin with searching for breather solution with a low-frequency mode, and $k$ is taken as $k = 2 \pi /L = \pi/4$. 
In terms of checking the validity of  Eq.~(\ref{cond_bound}), the amplitude $A$ and the squared speed of wave $\alpha$ are taken as free parameters.   
We have carried out systematics: 4$T \times$7$A \times$10$\alpha$ and the other random choices, which is up to $\sim$ 500 calculations.

The transient appearance and disappearance of breather solutions are distinguished by whether the mixture of plural numbers of mode are activated or not, and by whether the values of $u(t,x)$ for a given spatial interval keep the positivity or negativity.
For giving the criterion of choosing the values of $\alpha$, it is necessary to take sufficient numbers of $\alpha$ to identify the border between the appearance and disappearance of breather solutions.
Based on the bisection method, the border points are plotted if the relative error of the interval width is less than 0.10$\%$.
Here is a reason why we perform 10$\alpha$ times calculations for one combination of $T$ and $A$.

Even starting from exactly the same initial functions (Fig.~\ref{fig:ini}), some waves result in the simple oscillation (Fig.~\ref{fig-o}), and the breathing oscillation (Fig.~\ref{fig-b}) is achieved in the other cases.
Those difference is only in the difference of $\alpha$ value -0.220 and -0.075. 
In Fig.~\ref{fig:res1} the result for short time interval $T=64$ is shown.
The theoretical prediction (red curve) agrees quite well with the numerical systematics (black points).
Here we confirm the validity of the existence limit of breather solution for a given initial setting:
\begin{equation}
|\alpha| \le \frac{1}{k^2} \left(\mu-\frac{3}{8}\beta A^2 \right), 
\end{equation}
which is obtained by the polynomial regression of obtained border points.

Let us move on to the lifetime estimates of breather solution.
At points $x= 0,4,8$, the sufficiently small wave amplitude condition: $|u(x,t)| << 1$ is satisfied for both breathing and simply oscillating modes.
The first signal of instabilization of breathing and simply oscillating modes can be seen by observing the values at the fixed spatial points $x= 0,4,8$ whether $u(t,x)$ without satisfying $|u(x,t)| << 1$ appears or not.
We actually take $|u(x)| < 10^{-4}$ for the sufficiently small wave amplitude condition.
This criterion is exploited when we consider larger time intervals.
Note that all the calculations performed in this paper satisfies this smallness condition.
That is, we focus on the competitive existence between the breather solution and the simple oscillation.

Blue, green and purple curves in Fig.~\ref{fig:res2} are depicted in terms of a maximum value of parameter $- \alpha > 0$ of breathing modes to be stable within each time intervals $T = 128, 160, 192$. 
We see that, if we have longer time intervals, some breather solutions vanish depending on the values of $A$ and $\alpha$, and the ordinary oscillations appear instead.
In this sense, the value $T$ is regarded as the lifetime of breather solution.

\begin{figure}[tb]
  \begin{center}
  \hspace{-24mm}
  \includegraphics[width=50mm,bb=9 9 358 434]{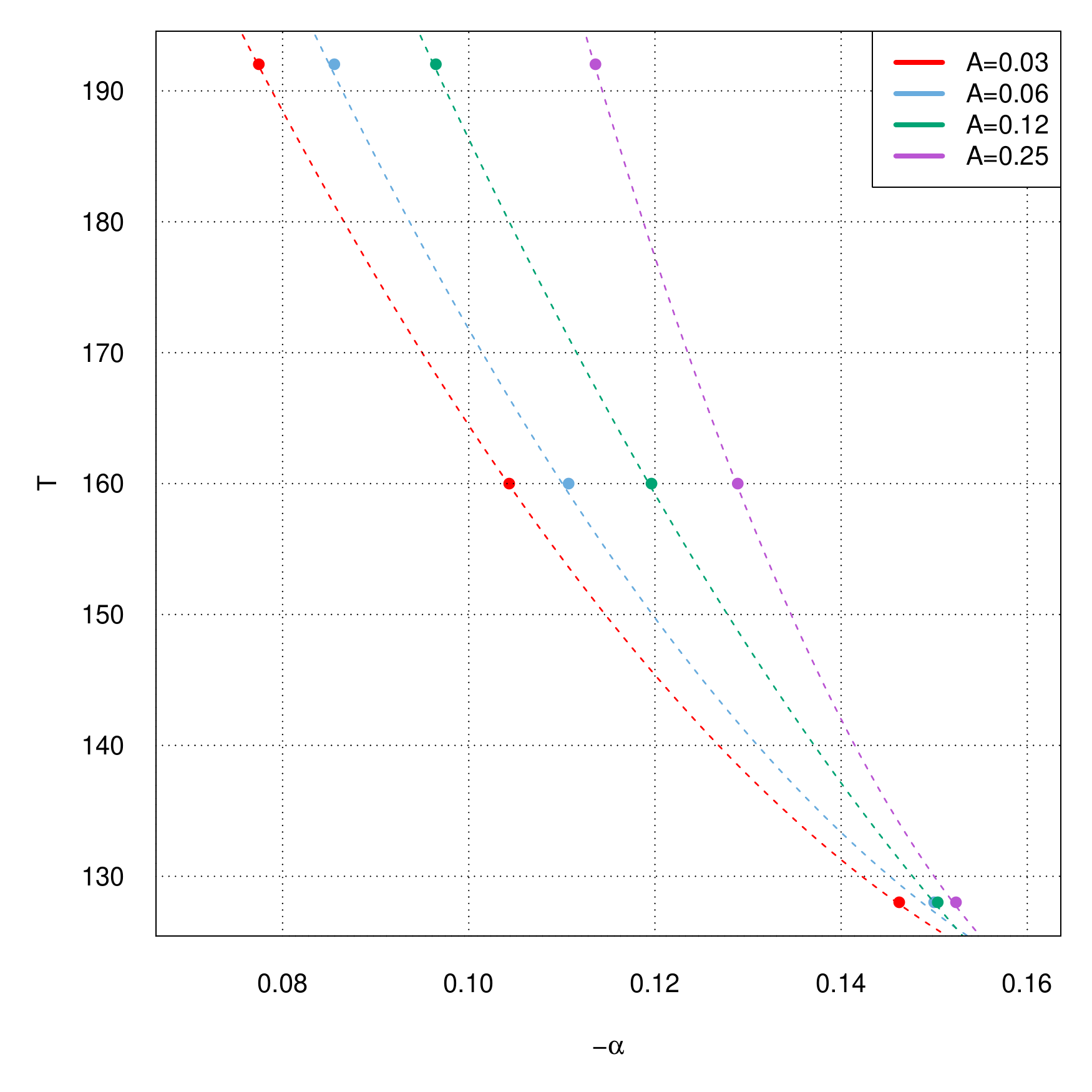}
  \caption{(Color online) Parameter dependence of lifetime $T$ of breather solution.
Parameters are fixed to $(\mu, \beta, L) = (0.125, 1, 8)$.
Comparison between four different ammplitude $A, (A = 0.03,\ 0.06,\ 0.12,\ 0.25)$ of initial function is shown in order to find the relationship between $T$ and $\alpha$.
The calculated values are shown by circles.
Red, blue, green, purple curves are obtained by the polynomial regression.
  }
\label{fig:ltime}
\end{center}
\end{figure}

Here, based on the numerical results, the relationship between the lifetime of the breather solution and the squared speed of wave $\alpha$ is shown in Fig.\ref{fig:ltime}.
As for the overall trend, it can be seen that the larger the squared speed of wave $\alpha$, the shorter the lifetime of the breather solution.
On the other hand, for a fixed the squared speed of wave $\alpha$, the larger the amplitude $A$ of the initial function results in the longer lifetime of the breather solution.
In conclusion the lifetime of the breather solution are determined only by the squared speed of wave $\alpha$ and the amplitude $A$ of the initial function.

\section{Conclusion}
In this paper, the space-time periodic breather solutions are numerically searched.
Since the periodic boundary condition is imposed, the obtained breather solution forms a closed curve in the phase space with respect both to time and space.
In this sense, what is meant by breather solution in this paper is closed compact manifold in the phase space.
From a different point of view, the breather solution is the localized oscillations around the constant distributions $u = \pm \sqrt {\mu/\beta}$.

The appearance condition of short- and long-lived breather solution is obtained in a purely theoretic mode analysis.
The validity of existence-limit formula  (\ref{cond_bound}) is supported by systematic numerical experiments with full nonlinearity. 
A mutual relation between coefficients is provided; the amplitude of wave must be smaller for the waves with higher speeds.
In the present settings, we see that the breather solution seems to decay asymptotically.
Consequently what we have obtained in this paper is the breather solutions with a finite lifetime.
\vspace{12mm} \\



\end{document}